# Enabling Secure and Usable Mobile Application: Revealing the Nuts and Bolts of software TPM in today's Mobile Devices

Ahmad-Atamli Reineh
University of Oxford
atamli@cs.ox.ac.uk

Giuseppe Petracca
Pennsylvania State University
gxp18@cse.psu.edu

Andrew Martin
University of Oxford
andrew.martin@cs.ox.ac.uk

Janne Uusilehto
Microsoft
janne.uusilehto@microsoft.com

*Abstract*— The emergence of mobile applications to execute sensitive operations has brought a myriad of security threats to both enterprises and users. In order to benefit from the large potential in smartphones there is a need to manage the risks arising from threats, while maintaining an easy interface for the users. In this paper we investigate the use of Trusted Platform Model (TPM) 2.0 to develop a secure application for smartphones using Windows Phone 8.1. In particular, we suggest a framework based on remote attestation as a proxy to authenticate remote services, where the device is associated to the user and replaces the user's credentials. In addition, we use the TPM 2.0 to enable secured information and data storage within the device itself. We present an implementation and performance evaluation of the suggested architecture that uses our novel attestation and authentication scheme and reveal the caveats of using software TPM in today's mobile devices.

## I. INTRODUCTION

Smartphones are being increasingly used to perform sensitive operations such as mobile banking transactions, storing sensitive data, and as a payment method. For applications to securely perform operations they have to rely mainly on the security provided by the operating system and third-party cryptographic libraries in order to design secure applications. The aforementioned give no choice to developers but to use and trust respectively the features offered by the OS provider and external security libraries provided by individuals. The latter leaves many developers limited in functionality to design an application that is secure and usable[1],[2], [3].

In order to secure operations, assure the integrity of the application, platform, and the software running on the device, trusted computing approaches have been adopted[2]–[6]. In many pieces of research, desktops and servers systems' were developed to assure the integrity and confidentiality of the code/data within an application. These systems use hardware elements, Trusted Platform Module (TPM) and Intel's Trusted Execution Technology (TXT).

Most of the proposed solutions are feasible on desktops and servers but not on mobile platforms and that's due to the absence of Intel TXT, Hardware TPM chip in such platforms. Alternatively, ARM based chips are being used in mobile platforms such as Nokia Lumia 830[7] and others. In which, ARM provides Trusted Execution Environment (TEE) called ARM TrustZone[8] that provides hardware memory protection for the trusted environment. The latter allows running smaller Trusted Computing Base (TCB) in an environment that is isolated from the rest of the execution.

In smartphones, several pieces of research used emulated TPM. However, these approaches used a software TPM either in the kernel level[9] or as an embedded library[10], but not in a trustworthy environment that is separated from other code in the system. Starting from Windows Phone 8.1, all Windows Phone devices include software TPM. The TPM provided by Microsoft, runs in trustworthy hardware, ARM TrustZone. In which, the TrustZone secures the execution of the TPM and separates such a sensitive component from the applications and most of the system.

Our contributions in this paper are:

1. We investigate the usage of TPM 2.0, and the ability of such component to enable useable security in mobile devices.

2. We present a prototype implementation of mobile banking application that uses TPM 2.0 and provide evaluation results for our proof of concept.

3. We present performance evaluation of a well-known TPM 2.0 functions that run in a mobile platform, Nokia Lumia 830.

4. We explain what is still missing, and what are the caveats in using software TPM in today's devices.

The rest of the paper is organised as follow. Section 2 describes related work on trusted computing in mobile devices. Section 3 explains features of trusted computing which are relevant to our work. In section 4 we propose a new framework for trusted mobile applications, making use of *remote attestation*. Section 5 describes a proof-of-concept implementation, and presents benchmarks for its operation. Section 6 evaluates the app framework using the TPM. The paper ends with an outline of what is missing when using software TPM, and some conclusions.

## II. RELATED WORK

The notion of using TPM for mobile phones is not new. However, there is no TPM chip available in mobile platforms. Several pieces of research adopted the TPM to preserve privacy and provide trust in the system.

Some approaches [9][11][12] used mini TPM emulator as a Linux Kernel Module (LKM) and is part of the android kernel. In [11], Nauman and colleges present an attestation mechanism for VM-based architecture OS, Android, and show the feasibility in terms of complexity and battery consumption.

In subsequent work by the same group [9], a protocol for keystroke dynamics analysis is proposed, in which it enables web-based applications to make use of remote attestation in android platform.

However, these approaches relay on an emulated TPM that is part of the android kernel, thus, the trustworthiness of this element is questionable. Such an approach does not get the benefits of a TPM being a secure element, which is protected from the OS and other code. Also, there is nothing mentioned about the execution environment and the TCB of the same environment. The TPM is good candidate that provides much functionality for securing a system. However, once compromised there is no guarantee that it can still assure the confidentiality of the data or the integrity of the system.

Other papers[13], [10], [14] extended the OS kernel to assure that a trusted code runs before the rest of the application, thus enables to verify the integrity before passing execution to applications in the same system. GuarDroid [10], provides a trusted path between the user and the app server by leveraging a service added to the OS. The user inserts his password to a trusted input provided by the OS, which in turn encrypted with a protected secret key. The point is to hide passwords from untrusted apps while preserving external behaviour. In [14], Zhang el al. develop a system to ensure that a secure kernel is booted. Once the kernel is booted, the work is done by the measured software to ensure a well-behaved execution of the system through an agent that enforces authenticated security policies.

These approaches assume trusted OS or other measured software that run in the same environment as the OS. The user is not protected from vulnerabilities in the code. Once the OS is compromised there is no assurance that the confidentiality of the data is kept.

Our approach is different in that it uses a TPM stand-alone from the rest of the system. We promote the use of a stand-alone element, the TPM, that is not an extension of the OS. We develop an application that makes use of software TPM that runs in a secure environment, ARM TrustZone. We show how a TPM can elevate security that is also usable in mobile platform.

## III. BACKGROUND

### A. Trusted computing

Trusted computing, presented by the Trusted Computing Group (TCG), introduced a unique security mechanism through hardware. The TPM from the family 2.0 [15] can be integrated into many different platforms including mobile devices and PCs. The TPM, a cryptographic co-processor, is designed to be resilient to software attacks and to possess unique security features for protecting secrets, randomness, sequencing, protected storage, and reporting [16]. It provides security functions like random number generator (RNG), *Platform Configuration Registers* (PCRs) that identify the platform, key generation, and various cryptographic functions. The TPM keeps track of the platform's state using the PCRs. The PCRs are divided into two parts; the static PCRs and the dynamic PCRs. The static registers take their initial value on TPM reset only (designed to coincide with platform reset) and are used for the evaluation during the boot stages to prevent access to secrets if the extended PCRs do not match the required data. They enable a chain of trust for the booting process.

The TCG introduced the concept of *Locality* that allows various trusted processes on the platform to communicate with the TPM such that the TPM is aware of which trusted process is sending commands. The second generation of TPM (TPM 2.0) includes a range of localities [17] to enable flexibility when using the dynamic PCRs. In particular, the TCG Mobile Phone Working Group (*MPWG*) introduces two new localities *(32 and 33) that are related to a Trusted Execution Environment (TEE)* [17]; locality 32 which indicates access from the code within the same TEE as the receiving TPM Mobile, and locality 33 which indicates access from an application TPM Mobile residing in the same TEE as a platform TPM mobile. The dynamic PCRs can be reset during run time in TPM2.0 using the TPM2_PCR_RESET when having the right permissions, giving more flexibility for the OS and the applications in managing the PCRs' values[15]. The authorisation level of the mobile applications and the operating system is determined according to the localities, where the operating system runs with locality value 2 and acts as the *Measured Launched Environment* (MLE) that manages the PCRs values. The application can use localities' values 32 and 34 in TEE as mentioned earlier, and the operating system acts as the dynamic Root of Trust for Measurements (RTM) and provides dynamic chain of trust during run time that starts from the MLE. The TPM can associate a certain data to a platform's state and identity using the sealing process. The TPM ensures association of the data to a platform by adding a nonce to the data package that associates the data package to an individual TPM. The record of the nonce is kept inside the TPM, and during the unsealing process the TPM validates the correctness of the nonce in the data package[16]. The PCRs are checked only during the unsealing process and only correct pre-selected values will lead to success of the operation.

*B. Remote attestation*

Remote attestation is a feature of the TPM intended to enable the affirmation of remote services. It reports PCR values using Attestation Identity keys (AIKs), allowing remote services to validate the content of the signature and the PCR values – a representation of some part of the platform state. The reported signature and PCRs are validated against a white list at the remote service, which is obtained on first registration to a service. The remote service uses the public part of the key to validate the AIK to the AIK credential. Through attestation, a platform can verify that only trusted software is executing, by verifying root of trust of the measured parts in the chain that lead to the executing software.

*C. Windows Phone 8.1 Security*

Windows Phone uses two latest standards-based security hardware components[18], UEFI and TPM, to keep the confidentiality of the data and the integrity of the platform and its software. UEFI ensures that the OS loader is secure, tamper free, or modified by an attacked. In addition, UEFI initializes the hardware and runs integrity checks that verify the integrity of the firmware before passing the executions. Such ability protects the firmware from rootkits and enables the transition from hardware to software after verifying its trustworthiness. The aforementioned enables booting the system securely and prevents attackers from jail breaking the device and installing uncertified apps.

UEFI and the OS use TPM to store integrity measurement, which verify that the measured software hasn't changed. Extending the hashes of the software using the TPM during the boot process brings the device to a trusted state, and any change in the software is reflected on the TPM which in turn refuses of decrypting and unsealing sensitive data within the device.

IV. FRAMEWORK FOR TRUSTED MOBILE APPLICATION

In this framework we use trusted computing technology, the TPM to secure sensitive data such as banking information. The Local Mobile Application (LMA) uses the TPM to store the identity and the credentials used for user authentication to the remote server. The local mobile application authenticates the user's identity in each access using a PIN or a fingerprint. The PIN is an important component for using the LMA; as it is used to access the application, in addition to being used as a configuration parameter for local TPM operations. The LMA establishes a trusted channel connection to the remote application server, and it undertakes mutual authentication with the bank server.

To focus on the strength of using TPM in a smartphone, we assume that the operating system is not compromised and each application runs in an isolated execution environment. In addition, we assume that a trusted code manages the dynamic PCRs when moving between applications by storing the PCRs value when another application is running and retrieves them when needed. We argue that these assumptions are reasonable in the context of mobile. Our framework consists of two parts, the *registration process* and the *login process* which both use a trusted channel to communicate between the device and the bank service. The channel carries all requests and responses between the user's device and the bank server through the secure TLS protocol connection to protect the exchanged data and authenticate the bank server. The bank maintains a white list of the acceptable entries which is obtained during the initial registration to the bank service: each entry is tied to a user's data and used by the bank for user authentication. The registration and login processes are explained in more detail in the following sections.

We use a sequence of authentication and attestation processes between the user, mobile application, and the bank service. The user authenticates to the mobile application and only then the app establishes a SSL/TLS session that authenticates the bank service, which then authenticates the user in two steps, and attests the mobile application. In the first authentication process, the activation key plays a major role in authenticating and associating the user with the app since the activation key is unique per user. This is important because only a certified app with appropriate activation key can authenticate to the bank service, thus, protecting the users against phishing attacks. In the second step, the app provides the user's credentials which are sent to the bank service to get access to the user's account. Both steps are essential in the initialisation process because the activation key is not associated to any user, and the credentials associate the activation key to a user.

*A. Mobile application registration*

**Fig. 1** describes the process of registering a new account with the mobile application. Applications can be downloaded from the App store of the platform provider: only signed and certified applications that have a unique license should be used. The license validity is verified by the OS before installing any app on the device. When starting the application the OS measures the software and extends the measurement into a dynamic PCR. We assume that the operating system uses application-specific PCRs and manages them according to the running application thus allowing us to use the dynamic PCR without worrying that other application will overwrite it. This gives control to the bank service, enabling it to revoke a configuration from its white list when a version is out of date or contains security holes. This method enables the bank to protect its users when discovering security vulnerability in the application they provide.

Upon first time start, an activation key is needed to unlock the app: this is provided to the user by the bank. It is used to associate an application to the user and allows stronger authentication with every access. The activation key is hashed and extended into the dynamic PCRs, and stored safely using a storage key (SK) created by the TPM.

Using the TPM to verify the user's identity mitigates software attacks, while keeping the integrity and confidentiality of the user. Using a dedicated key, fingerprint or password (e.g. PIN), the user's identity is proved to the application at every use. This key is hashed by the TPM and secured using a SK which associates the user's key to the platform. The application verifies the authenticity of the user's key with every start/resume of the application.

The local mobile application (LMA) establishes a trusted channel with the remote application using TLS to enable secure communication between the mobile and the bank server. The application requests the account details in order to access it from the remote application, which in turn authenticates the user's details and creates an access certificate for future access. The application stores these results in the TPM once and uses them as OpenID to get the data for the requesting service when working with two step authentication. A blob with the user's credentials, associated to the relevant dynamic PCRs is created and stored in the device. The dynamic PCR state is signed using an AIK and sent to the bank for user attestation with every connection.

### B. Secure mobile login

**Fig. 2** describes the sequence of events in using the app after completing the registration process. In order to establish a connection with the bank, it is important to attest and authenticate the mobile application and the user's data stored within it against the bank's records. The OS measures the application and extends the value into the dynamic PCR, which then passes the control to the banking application. The app requests a PIN for user authentication, and it uses the TPM to hash the PIN before extending it to the dynamic PCRs. In order to communicate with the bank server, the device needs to attest and authenticate its identity. The app extracts the user's data through TPM unseal operation on the stored blob; the success of the operation requires the right values of the dynamic PCRs. The three conditions that need to be met for successful unsealing operation are right (1) values for the activation key, (2) hashed software, and (3) user PIN. These PCR values reflect the state of the device; they are signed using AIK and sent to the bank server to complete the client's attestation and authentication between the bank and the user's device, for establishing a trusted channel. The bank authenticates the user's device PCR values against a list of known "good values"[19]. On completion of the attestation process, the user's credentials are sent automatically by the mobile device to the bank server for authentication. The bank verifies the user's data and grants access to the user. The user's credentials are used for first time accessing the bank services and on later access the two-step authentication provides two independent pieces of authenticity.

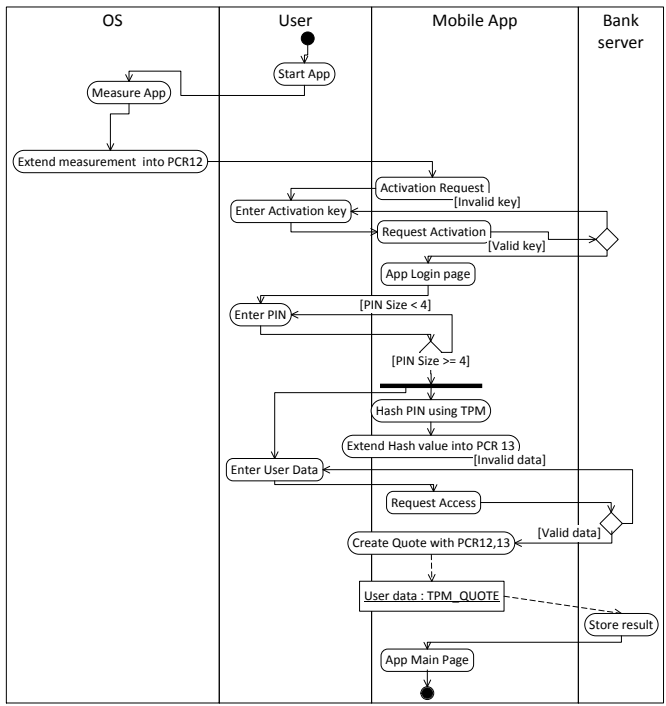

Fig. 1. User Registration Flow

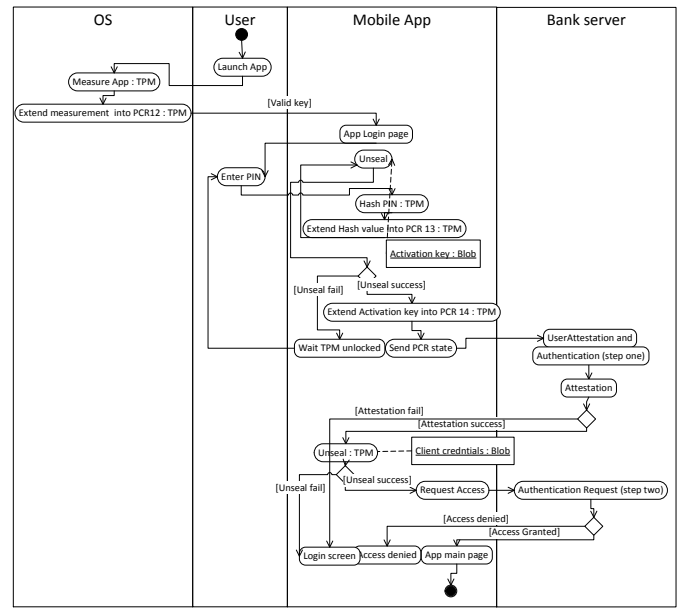

Fig. 2. User Login Flow

## V. PROOF OF CONCEPT

In this section we describe the implementation of the application's architecture described in the previous sections. The objective of this implementation is to demonstrate that using current technologies, it is realisable to achieve the security standards we specified.

### A. Implementation Specifications

We implement an application for Windows Phone 8.1. Windows Phone 8.1 based architecture is Windows NT kernel, the same kernel used for Windows 8. ".Net" can be used for developing applications for both operating systems and the implementation used for developing an app in one environment, can be suitable for the other environment. We designed the application for Windows Phone 8.1. We use C# to develop an elementary application that uses the software TPM in a smartphone. The application communicates with TPM2.0 through Windows TBS (TPM Base Services) system call in C#. In order to use the TPM we use TBS function that receives "TPM command" in binary format as an input to the TPM.

In our implementation we couldn't use the TSS.Net library[20] since it is not compatible with Windows phone environment. Notwithstanding, in the near future these libraries should be able to support Windows phone environment since the base code, as introduced by the TCG specification for communicating with the TPM, is in C.

Our system consists of two devices and a channel to communicate between them. The first runs the bank application (the client) and the second runs the bank services. The communication channel between the two ends uses SSL protocol for communication, and we use the SSL library developed by Microsoft to make use of this protocol.

### B. TSS.Net

The TSS.Net library is a managed code written in C#. The library is open source and is used for easy access to the TPM. It makes it easier writing Windows application using the TPM. The TSS.Net handles all the low level functionality when using the TPM, hiding all the complex interactions from the developer. In addition, the library communicates with the TPM simulator through a network socket TCP/IP to give the opportunity of development and debugging.

### C. TPM2.0

The TPM2.0 simulator was developed by Microsoft and the TCG based on the TPM 2.0 specification [15]. It emulates the TPM 2.0 commands in software. To our knowledge there is no hardware TPM aligned to TPM 2.0 specification. We use the TPM 2.0 simulator as a tool to test our framework. Microsoft surface [21], and Nokia Lumia 830 include a TPM 2.0 simulator running in the TEE of Arm TrustZone[8] processor to protect its secrets.

## VI. EVALUATION

The goal of our design is to provide a framework for mobile applications which is easy to use and at the same time does not compromise security. In order for users to access their accounts, they are required only to authenticate themselves to the mobile application by providing a PIN of their choice. The initialisation process requires more efforts from the user by requesting all the relevant data such as activation code of the application, gating PIN to the application, and user credentials to access the bank services. This information is initially obtained from the bank. However, it is performed for one instance only during the setup time and the user is exempt from providing these data with every future access to the bank account.

In order to provide a secure end-to-end solution we recommend trusted computing technology, the TPM to protect the integrity and confidentiality of the user's data, SSL for secure channel connection, and two level authentications. The first level authentication in the login process can be done with the attestation data only, this association between the authentication and attestation in the setup process enables the user to replace bank credentials with attestation data for access control to the bank services. The use of these security elements are hidden from the user making it easier to use, and at the same time provides secure connection to the bank services.

### A. System benchmarks

The proof of concept implementation was tested on a Nokia Lumia 830[7] with an Arm Quad-Core Cortex-A7 CPU running at a clock frequency of 1.2 GHz and TPM2 simulator (version 1.0) running in the secure world of Arm TrustZone. Using our implementation we are able to evaluate the time scale and relative time costs for the different operation performed.

Table 1 shows the time measurements of the operation performed on a "software TPM 2.0" that runs in the secure zone of ARM TrustZone processor. In order to test the performance of these operations in Nokia Lumia 830, we executed each operation 10000 and assume equal probability for each measurement to calculate the Standard Deviation (Std Dev).

Table 3 shows the differences between a credential-approach compared to the several stages we suggested in our framework. We determined the credentials' size based on the banking application. Our initialisation process uses more data compared to the credentials approach. However, in steady state we use less data when working with one step authentication and more data when working with two steps. The two step authentication adds another level of security where the bank checks the user credentials with every access by the user, where the one step uses the remote attestation and application Activation Key only.

**Table 1.** Time of execution measurements.

| Operation | Time (ms) | [Std Dev] |
|---|---|---|
| RNG | 0.896 | [0.17] |
| PCR Read | 1.05 | [0.12] |
| Sha1 Data Hash | 1.06 | [0.15] |
| Sha1 Key Sign | 0.4 | [0.07] |
| Extend PCR | 1.2 | [0.11] |

**Table 3.** Number of bytes used in the two approaches

| | Bytes |
|---|---|
| Credential based | 44 |
| Registration process | 84 |
| Login (1 Step) | 40 |
| Login (2 Steps) | 84 |

Table 4 summaries the number of operations used in the registration and the log-in processes respectively. In the login process, one unsealing operation could be spared when giving up the second authentication stage.

**Table 4.** The Number of operations in the Registration and login process

| Operation | Number of calls (Login process) | Number of calls (Registration process) |
|---|---|---|
| Sealing | 0 | 2 |
| Hashing | 1 | 1 |
| Unsealing | 1-2 | 0 |
| AIK Generation | 1 | 1 |
| Extend PCR | 3 | 3 |

Table 5 summarises the points addressed in this paper compared to existing technology. We mention how current technology addresses each point, the flaws that exist in current technology, and how does our framework overcome them.

**Table 5.** Mobile Application summary compared to current technology

| Points | Current solution | Vulnerability/issues | Our solution |
|---|---|---|---|
| Brute force attack | External device password generator | Requires additional device | TPM anti-hammering mechanism |
| Man in the middle | SSL/TLS | Certificates managing and handling | Dual authentication and remote attestation |

## VII. WHAT IS MISSING?

In modern mobile operating systems e.g. Android, iOS, and Windows Phone, the OS enforces permission based access control to secure an element such as a file, a resource, and a service. This approach is proving to be inefficient as the number of vulnerabilities in OS is rising.

The use of software TPM allows the use of standards-based security that is also flexible to adapt in several systems such as smartphones and cloud environment. However, it is important to take into consideration the attack vector that rises in a system from adopting such an approach. The approach in smartphones for using software TPM is running such sensitive element in a secure environment that is isolated from the applications, Arm TrustZone. This environment is referred to as the *secure world*, and it's protected from applications that run in the *non-secure world*. However, the software TPM is not the only code executing in the *secure world* and there are others such as part of the OS and the kernel that run in such environment. Hence, the TPM is not protected from vulnerabilities in the rest of the code that has access to the *secure world*, e.g. the kernel. In the aforementioned approach, the TPM is a stand-alone element from the OS and the kernel, and runs in a trusted secure environment. However, it cannot be considered as a secure element as a hardware TPM.

### A. Secure Element

While the integrity of the software can be checked through verifying the hashed value at boot time, it does not guarantee the integrity of the code during run-time. A change in the execution can be easily achieved if other software has access to the executing environment. To counter this shortcoming, isolating an element from the rest of the system can provide a smaller TCB that is limited to the code of the desired element. A *secure element (SE)* is a combination of hardware and software and it provides us with secure environment of execution that is protected from the rest of the system. Nowadays, many mobile devices possess ARM processors that provide TEE. The TEE by itself does not provide the capability to create a SE since the TCB is not limited to software TPM but also to part of the OS.

### B. Trusted Path

Trust path refers to assuring that the user's input gets to the desired destination only and is protected from an untrusted app or the OS. The TPM can protect against brute force attack of passwords/PINs using anti-hammering, according to which, the TPM locks itself after configurable number of wrong attempts. This prevents the attacker from guessing the password in a reasonable time. However, passwords can be sniffed by untrusted application and the OS. This can be used by an untrusted app to access sensitive information and authenticate itself to the TPM and perform any operation like a valid user.

## VIII. CONCLUSIONS

Mobile applications have gained popularity with the advancement of mobile technologies. Many approaches have been suggested to solve security issues that arise when using mobile devices. Many papers addressing security in mobile applications rely on old technologies and suggest solutions within their limitations. However, many of these methods [22] are point solutions and are not easy to use, which present a serious limitation since the user becomes discouraged by the many layers of security measures.

Many methods aim to enhance security of mobile applications, by requiring more user input like copying codes from a smart card or SMS message, putting more responsibility on the users to enhance security. These methods became impractical and inconvenient to use, and having simple access control while not compromising security becomes essential. The use of a PIN is an easy access control for mobile application and hides all the security levels from the user making it very convenient and simple to use.

In this paper we show how using TPM 2.0 mitigate some of the existing threats on mobile applications using Trusted Computing Technology, the TPM, to improve current security schemes for mobile applications (Table 5). We propose a novel communication protocol that uses both authentication and attestation for setting up the user's account, hence removing the pressure of securing the credential by the user alone. In addition, we point to the missing pieces in using this technology. Our prototype implementation demonstrates feasibility and a path to an implementation on a mobile platform with the necessary trust characteristics.

*Acknowledgments. The authors would like to thank Andrew Paverd and Janne Uusilehto for many fruitful discussions.*